\documentclass{ws-mpla}
\usepackage{graphicx}
\usepackage[super,compress]{cite}
\usepackage{hyperref}
\hypersetup{colorlinks,urlcolor=black,citecolor=black,linkcolor=black,filecolor=black}
\DeclareMathOperator{\sgn}{sgn}

\begin{document}

\markboth{Alina Sagaydak and Zurab Silagadze}{On Finslerian extension of special relativity}

\catchline{}{}{}{}{}

\title{On Finslerian extension of special relativity}

\author{Alina E. Sagaydak}  

\address{Novosibirsk State University, 630 090, Novosibirsk, Russia. \\ a.sagaidak@g.nsu.ru} 

\author{Zurab K. Silagadze} 
\address{Budker Institute of Nuclear Physics and Novosibirsk State University, 630 090, Novosibirsk, Russia. \\ silagadze@inp.nsk.su}

\maketitle

\pub{Received (Day Month Year)}{Revised (Day Month Year)}

\begin{abstract}
We demonstrate that Robb-Geroch's definition of a relativistic interval admits a simple and fairly natural generalization leading to a Finsler extension of special relativity. Another justification for such an extension goes back to the works of Lalan and Alway and, finally, was put on a solid basis and systematically investigated by Bogoslovsky under the name "Special-relativistic theory of locally anisotropic space-time". The isometry group of this space-time, $\mathrm{DISIM}_b(2)$, is a deformation of the Cohen and Glashow’s very special relativity symmetry group $\mathrm{ISIM}(2)$. Thus, the deformation parameter $b$ can be regarded as an analog of the cosmological constant characterizing the deformation of the Poincar\'{e} group into the de Sitter (anti-de Sitter) group. The simplicity and naturalness of Finslerian extension in the context of this article adds weight to the argument that the possibility of a nonzero value of $b$ should be carefully considered. 

\end{abstract}

\section{Introduction}
Historically, special relativity arose as a result of attempts to resolve the conflict between Maxwell's equations and Galilean transformations \cite{Zahar}, resulting in the realization that the Lorentzian symmetry of Maxwell's equations is inherent in all physical phenomena. 
However, the symmetry group of Maxwell’s equations is the 15-parameter conformal group \cite{Cunningham,Bateman}, while only its 10-parameter subgroup, the Poincar\'{e} group, is recognized as a symmetry group of nature in the absence of gravity (space-time curvature).

Although Lorentz invariance has been thoroughly tested \cite{Zhang,Mattingly},  many of the ways in which Lorentz symmetry could break at very high energies, near the Planck scale, as predicted in some models of quantum gravity and string theory, remain untested (see, for example, \cite{Tasson:2014dfa,Kostelecky:1988zi,Blas:2010hb,Amelino-Camelia:2008aez,Liberati:2009pf,Liberati:2013xla}).

One particularly interesting and subtle suggestion about how the Lorentz symmetry could be broken was made by Cohen and Glashow \cite{Cohen:2006ky}. In the resulting theory, which the authors call {\it very special relativity}, Lorentz symmetry is broken very mildly if we assume that the true symmetry group of nature is not the Lorentz group, but its proper subgroup $\mathrm{SIM(2)}$. The inclusion of any discrete symmetries $P$, $T$ or $CP$ enlarges $\mathrm{SIM(2)}$ to the full Lorentz group. As a result, Lorentz-violating effects in very special relativity are absent for electromagnetic and strong interactions, as well as for any other theories that preserve any of these discrete symmetries. Since the effects of $CP$ violation in weak interactions are small, it can be expected that the effects of Lorentz violation in very special relativity will necessarily be small \cite{Cohen:2006ky}.

According to Segal's heuristic principle, only such mathematical structures that are stable against small deformations are realized in ultimate physical theories \cite{Segal,Finkelstein:2005mc} (see also \cite{VilelaMendes:1994zg,Faddeev1998}). If the algebraic structure of the physical theory is not stable against small deformations, then this indicates that the physical theory can be generalized to a wider class of physical phenomena. 
For example, quantum mechanics can be viewed as a deformation of classical mechanics, with the Planck constant playing the role of the deformation parameter, and the transition from non-relativistic to relativistic dynamics is associated with the deformation of the Galileo group into the Lorentz group, where the speed of light is the deformation parameter  \cite{VilelaMendes:1994zg,Faddeev1998}.

The Lorentz algebra, being semi-simple, is stable against small deformations. However, if we add space-time translations, we get a Poincar\'{e} algebra, which is again unstable and can be deformed into its stabilized version, which is the de Sitter (or anti-de Sitter) algebra \cite{Bacry:1968zf,Dyson:1972sd}. Thus, it is probably not surprising that observations show that the cosmological constant is not zero \cite {Supernova,Planck:2018vyg}, albeit mysteriously small \cite{Peebles:2002gy,Carroll:2000fy}.

Like the Poincar\'{e} group, its subgroup $\mathrm{ISIM(2)}$ ($\mathrm{SIM(2)}$, supplemented by space-time translations) is subject to deformations into a 1-parameter family of groups $\mathrm{DISIM}_b(2)$ \cite{Gibbons:2007iu}. For any values of $b$, $\mathrm{DISIM}_b(2)$ is an 8-parameter subgroup of the 11-parameter Weyl group (the semi-direct product of dilatations with the Poincar\'{e} group). The group $\mathrm{DISIM}_b(2)$ does not leave the Minkowski metric $\eta_{\mu\nu}dx^\mu dx^\nu$ invariant. Instead, this deformation of $\mathrm{ISIM (2)}$ naturally leads to a pseudo-Finsler geometry with the metric \cite{Gibbons:2007iu}
\begin{equation}
ds^2=(n_\lambda dx^\lambda)^{2b}(\eta_{\mu\nu}dx^\mu dx^\nu)^{1-b}, 
\label{eq1}
\end{equation}
where $n^\mu$ is a fixed null vector. Therefore, according to Segal's principle, we can expect that $b$ is not zero, but it can be extremely small by analogy with the cosmological constant \cite{Dhasmana:2019jil}. 

The unit vector $\vec{n}$ in $n^\mu=(1,\vec{n})$ indicates the preferred direction in three-dimensional space. Thus, in this case, the space is not isotropic. Note, however, that (\ref{eq1}) preserves the conformal structure (light cones) of Minkowski space and, therefore, the speed of light does not depend on the direction of its propagation (and is equal to one). 

It was Lalan who was the first to understand (in two-dimensional case) that, having discarded the hypothesis of isotropic space, space-time can be represented as a pseudo-Finsler space, and not as a pseudo-Riemannian manifold \cite{Lalan}.

The anisotropy introduced by the pseudo-Finsler metric (\ref{eq1}) is physical and must be distinguished from the anisotropy of light propagation introduced by any non-standard synchronization procedure. The latter does not change the character of the Minkowski space-time, and in this case the anisotropy is simply an artifact of the use of non-Lorentzian coordinates to describe the Minkowski metric \cite{Sonego:2008iu}.

The four-dimensional pseudo-Finsler metric (\ref{eq1}) first appeared in \cite{Alway} (without citing Lalan), but this publication did not attract attention. Soon the possibility of Finslerian extension of relativity was rediscovered by Bogoslovsky \cite{Bogoslovsky:1973zen,Bogoslovsky}, and then thoroughly investigated (see \cite{Bogoslovsky:1994cq,Bogoslovsky1992,Bogoslovsky:1999pp,Bogoslovsky:1993vu} and references therein).

In this article, we demonstrate that the pseudo-Finsler metric (\ref{eq1}) is a very natural generalization of the usual relativistic interval, and therefore the possibility of non-zero $b$ should be seriously considered. Although experimental evidence  already severely  constrains $b$, and it is tempting to repeat the old conclusion that such a generalization of special relativity is irrelevant \cite{Strand}, in reality the question of "why is $b$ so small" is just as fundamental as the mystery of the cosmological constant \cite{Gibbons:2007iu}, and therefore deserves the same careful study. 

\section{Robb-Geroch's definition of relativistic interval}
Light cones define a partial conical order in space-time: for a given event at the apex of the light cone, we have absolute concepts of time relations {\it before} (events in the backward light cone) and {\it after} (events in the forward light cone) \cite{Robb1,Robb2,Robb3,Robb4}. 

The concept of an ideal clock that measures a proper time along its time-like world line makes it possible to quantify temporal relationships, but only along a given world line. To extend these temporal relations beyond the world line of an ideal clock, we need the concept of {\it simultaneity}. However, there is no absolute concept of simultaneity in space-time, and we must define it by convention (by a stipulation, as Einstein put it) \cite{Einstein:1905ve}. 

The definition of simultaneity, introduced in Einstein's seminal paper \cite{Einstein:1905ve}, was actually previously considered by Poincar\'{e} (see \cite{Chashchina:2016tey} and references therein. Essentially the same operational definition of distant simultaneity was advocated by St. Augustine in his {\it Confessions}, written in AD 397 \cite{Jammer}). However, Poincar\'{e} never attributed a crucial significance to this definition of simultaneity \cite{Galison}. Thus, we will not sin too much against historical truth if we call standard synchrony simply Einstein's simultaneity. 

Einstein's simultaneity is defined as follows. If a light pulse is sent from a time-like curve $\gamma$ to some nearby event $A$ at the proper time $\tau_1$ on $\gamma$, being immediately reflected back at $A$ and received on $\gamma$ at the proper time $\tau_2$,  then an event $B$ on $\gamma$ with the proper time $\tau=\frac{1}{2}(\tau_1+\tau_2)$ will be simultaneous with $A$ according to Einstein. The distance between simultaneous events $A$ and $B$ is defined as $l=\frac{1}{2}(\tau_2-\tau_1)$ \cite{Robb4}. In the usual space and time picture, these definitions are equivalent to Einstein's two postulates, provided that the units of temporal and spatial measurements are chosen so that the speed of light is equal to one.

Since $\tau=\frac{1}{2}(\tau_1+\tau_2)$, the distance $l$ can be written in the form
\begin{equation}
l^2=(\tau_2-\tau)(\tau-\tau_1).
\label{eq2}
\end{equation}
This form makes it possible to extend the concept of "distance" to all sufficiently close events, not necessarily simultaneous, and to introduce the concept of a relativistic interval \cite{Geroch1,Geroch2} (see also \cite{Synge,Salecker:1957be})
\begin{equation}
s^2=(\tau_2-\tau)(\tau-\tau_1).
\label{eq3}
\end{equation}
The proper time is determined only along the world line of the particle (observer). The concept of simultaneity allows us to extend the concept of time to events in the immediate vicinity of some referential event. In particular, let an event $O$ be the referential event corresponding to the proper time $\tau=0$ for the world line of an observer $\gamma$. The tangent to the world line of the observer $\gamma$ at $O$ defines the natural time direction for this observer. Then the so-called radar coordinates $(t,x)$ (for simplicity, we consider mostly two-dimensional space-time, so beloved by philosophers and researchers of the foundations of special relativity) are defined as follows \cite{Robb4,Synge1921}. For any event $A$ in an infinitesimal neighborhood of $O$, the time coordinate $t=\tau$ is equal to the proper time $\tau$ of the event $B$ on the world line $\gamma$, which is simultaneous with $A$. The $x$ coordinate is the distance from $A$ to $B$. Therefore,
\begin{equation}
t=\frac{1}{2}(\tau_1+\tau_2),\;\;\;x=\frac{1}{2}(\tau_2-\tau_1),
\label{eq4}
\end{equation}
where the meanings of $\tau_1$ and $\tau_2$ were described earlier.

In general space-times, these definitions are local in the sense that the indicated procedure with light signals in general works only for sufficiently close events \cite{Perlick:2007np}. However, in flat (Minkowski) space-time and for inertial observers (whose world lines are straight lines), the radar coordinates can be extended to all space-time, and they will determine the division (foliation) of space-time into private time $t$ and the private space of the observer $S$ with world line $\gamma$. 

Observers whose world lines are parallel to the world line of $S$ share the same foliation of Minkowski space-time. The congruence of such world lines constitutes the inertial frame of reference, which we also denote by $S$. One can say that in special relativity, the inertial frame of reference is a set of freely floating ideal clocks with parallel world lines filling all space-time \cite{Schild}. 

In the inertial reference frame $S$, the private time and private space of the observer $S$ are elevated to the status of public time and public space of the inertial frame $S$. In general, public time and public space depend on the congruence of the world lines of observers sharing these concepts, and they differ from the private time and private space of a particular observer from this congruence (see \cite{Milne,Chashchina:2014gsa}, from which we borrowed the terminology).

In the following, we consider only inertial reference frames in Minkowski space-time and call the radar coordinates simply coordinates. 

\section{Finslerian generalization}
In (\ref{eq2}), $\tau_2-\tau=\tau-\tau_1$. Therefore, the distance $l$ will not change, if we define it as follows
\begin{equation}
l^2=\left(\frac{\tau-\tau_1}{\tau_2-\tau}\right )^b (\tau_2-\tau)(\tau-\tau_1),
\label{eq5}
\end{equation}
where $b$ is some real parameter. However, if (\ref{eq5}) is elevated to the definition of the relativistic interval
\begin{equation}
s^2=\left | \frac{\tau-\tau_1}{\tau_2-\tau}\right |^b (\tau_2-\tau)(\tau-\tau_1)=
|\tau_2-\tau|^{1-b}\,|\tau-\tau_1|^{1+b},
\label{eq6}
\end{equation}
then for $\tau_2-\tau\ne\tau-\tau_1$ we get a completely different interval, which, however, preserves the structure of the light cone if $b<1$ (in fact, we assume $b\ll 1$ to have only a slight deformation of the Minkowski space-time).

The modification (\ref{eq6}) of the relativistic interval may seem like an ad hoc choice. In fact, this generalization is essentially unique. Indeed, one might think that instead of (\ref{eq6}) one can take a more general expression
\begin{equation}
s^2=f\left (\frac{\tau-\tau_1}{\tau_2-\tau}\right )\,(\tau_2-\tau)(\tau-\tau_1),
\label{eq27}
\end{equation}
where $f(x)$ is some function of the dimensionless quantity $x=(\tau-\tau_1)/(\tau_2-\tau)$ with the property $f(1)=1$. However, the requirement that the infinitesimal interval does not depend on the world line $\gamma$, used in the definition of the Robb-Geroch interval, severely restricts the admissible form of the function $f(x)$. For simplicity of presentation, we consider this requirement in the case (\ref{eq6}) and comment on the general case at the end.
\begin{figure}[ht]
\begin{center}
\includegraphics[width=0.4\textwidth]{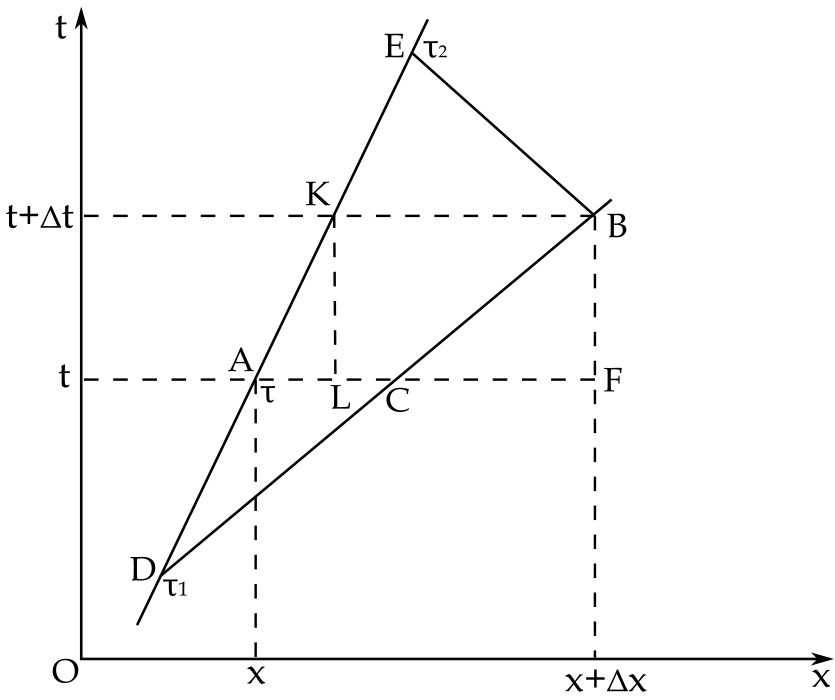}
\end{center}
\caption{Robb-Geroch's method of calculating relativistic interval between infinitesimally close events.}
\label{fig1}
\end{figure}

In the reference frame $S$, let us consider two infinitesimally close events $A$ and $B$ with coordinates $(t,x)$ and $(t+\Delta t, x+\Delta x)$. Our goal is to express the squared interval $ds^2$ between these events through coordinate differences $\Delta t$ and $\Delta x$. Let $DAKE$ be some world line passing through the event $A$ (see Fig.\ref{fig1}. Since we are considering infinitesimally close events, we can assume that $DAKE$ is a segment of a straight line). If the angle of inclination of the world line $DAKE$ to the $x$ axis is equal to $\theta$, then
\begin{equation}
\tan{\theta}=\frac{1}{\beta},\;\;\;\sin{\theta}=\frac{1}{\sqrt{1+\beta^2}},\;\;\;\cos{\theta}=\frac{\beta}{\sqrt{1+\beta^2}},
\label{eq7}
\end{equation}
where $\beta$ is the velocity corresponding to the world line $DAKE$ in coordinates $t,x$. The photon world line $DB$ is inclined at an angle $\pi/4$. Therefore, $CF=BF=\Delta t$. In the triangle $ADC$, $\angle CDA=\theta-\pi/4$, $\angle ACD=\pi/4$. Therefore,
\begin{equation}
\frac{AD}{\sin{\frac{\pi}{4}}}=\frac{AC}{\sin{\left (\theta-\frac{\pi}{4}\right )}},   
\label{eq8}
\end{equation}
and
\begin{equation}
AD=AC\,\frac{\sin{\frac{\pi}{4}}}{\sin{\left (\theta-\frac{\pi}{4}\right )}}=\frac{\Delta x-\Delta t}{\sin{\theta}-\cos{\theta}}=\frac{\sqrt{1+\beta^2}}{1-\beta}\,\left (\Delta x-\Delta t\right ).
\label{eq9}
\end{equation}
It follows from this result that
\begin{equation}
\tau-\tau_1=k(\beta)\,AD=k(\beta)\,\frac{\sqrt{1+\beta^2}}{1-\beta}\,\left (\Delta x-\Delta t\right ),   
\label{eq10}
\end{equation}
where $k(\beta)$ is a coefficient that transforms the Euclidean length of the segment $AD$ into proper time.

Likewise, we can calculate $\tau_2-\tau$, guided again by Fig.1. We have $KL=\Delta t$, $AL=KL\,\cot{\theta}=\beta\Delta t$ and $KB=\Delta x-AL=\Delta x-\beta\Delta t$. Since $\angle EKB =\theta$ and $\angle KBE =\pi/4$, the theorem of sines in the triangle $KEB$ gives
\begin{equation}
\frac{KE}{\sin{\frac{\pi}{4}}}=\frac{KB}{\sin{\left(\pi-\theta-\frac{\pi}{4}\right )}}=\frac{\Delta x-\beta\Delta t}{\sin{\left (\theta+\frac{\pi}{4}\right )}}.
\label{eq11}
\end{equation}
Therefore,
\begin{equation}
KE=\left (\Delta x-\beta\Delta t\right ) \frac { \sin { \frac{\pi} {4}}}{\sin{\left (\theta+\frac{\pi}{4}\right )}}=\frac{\Delta x-\beta\Delta t}{\cos{\theta}+\sin{\theta}}=\frac{\sqrt{1+\beta^2}}{1+\beta}\,\left (\Delta x-\beta\Delta t\right ).
\label{eq12}
\end{equation}
But $AE=AK+KE=KL/\sin{\theta}+KE$. Therefore,
\begin{equation}
AE=\sqrt{1+\beta^2}\,\Delta t+\frac{\sqrt{1+\beta^2}}{1+\beta}\,\left (\Delta x-\beta\Delta t\right )=\frac{\sqrt{1+\beta^2}}{1+\beta}\,\left (\Delta x+\Delta t\right ),
\label{eq13}
\end{equation}
and
\begin{equation}
\tau_2-\tau=k(\beta)\,AE=k(\beta)\,\frac{\sqrt{1+\beta^2}}{1+\beta}\,\left (\Delta x+\Delta t\right ).
\label{eq14}
\end{equation}
Then it follows from (\ref{eq6}), (\ref{eq10}) and (\ref{eq14}) that
\begin{equation}
\begin{aligned}
ds^2=&k^2(\beta)\left(\frac{1+\beta}{1-\beta}\right)^b\frac{1+\beta^2}{1-\beta^2}\left|\frac{\Delta x-\Delta t}{\Delta x+\Delta t}\right|^b(\Delta x^2-\Delta t^2)=\\
&k^2(\beta)\left(\frac{1+\beta}{1-\beta}\right)^b\frac{1+\beta^2}{1-\beta^2}\left (\Delta x -\Delta t\right)^{2b}\left |\Delta x^2-\Delta t^2\right |^{1-b}.
\end{aligned}
\label{eq15}
\end{equation}
The interval $ds^2$ between the events $A$ and $B$ will not depend on the world line $DAKE$ used in its calculation, if the $\beta$-dependent factor in (\ref{eq15}) is in fact a constant. The requirement $k(0)=1$ fixes this constant to be one, and we get
\begin{equation}
k^2(\beta)\left(\frac{1+\beta}{1-\beta}\right)^b\frac{1+\beta^2}{1-\beta^2}=1,\;\;\;
k(\beta)=\sqrt{\frac{1-\beta^2}{1+\beta^2}}\left (\frac{1-\beta}{1+\beta}\right )^{b/2}.
\label{eq16}
\end{equation}
Note that the resulting interval is exactly of the Finslerian type (\ref{eq1}).

In the general case (\ref{eq27}), we get instead of (\ref{eq15})
\begin{equation}
ds^2=f\left(\frac{1+\beta}{1-\beta}\,\frac{\Delta x-\Delta t}{\Delta x+\Delta t}\right )k(\beta)^2\,\frac{1+\beta^2}{1-\beta^2}\,
(\Delta x^2-\Delta t^2).
\label{eq27A}
\end{equation}
This expression will not depend on $\beta$ if 
\begin{equation}
k(\beta)=\left [\frac{1+\beta^2}{1-\beta^2}\,f\left (\frac{1+\beta}{1-\beta}\right )\right ]^{-1/2},
\label{eq30}
\end{equation}
and $f(\beta)$ is a multiplicative function. That is $f(1)=1$ and for all real $x$, $y$ we have
\begin{equation}
f(xy)=f(x)f(y).
\label{eq28}
\end{equation}
In this case we end up with
\begin{equation}
ds^2=f\left (\frac{\Delta x-\Delta t}{\Delta x+\Delta t}\right )\,(\Delta x^2-\Delta t^2).
\label{eq29}
\end{equation}
However, it is well known that the only non-constant and continuous (except, possibly, at the point $x=0$) solutions of the functional equation (\ref{eq28}) are \cite{Aczel}
\begin{equation}
f(x)=\left \{\begin{array}{l} |x|^b,\;\;\mathrm{if}\;\;x\ne 0, \\ \\
0,\;\;\;\;\mathrm{if}\;\;x=0,\end{array}\right . \;\;\;\mathrm{and}\;\;\;
f(x)=\left \{\begin{array}{l} |x|^b\sgn{x},\;b\ne 0,\;\;\;\;\mathrm{if}\;\;x\ne 0, \\ \\
0,\quad\quad\mathrm{if}\;\;x=0.\end{array}\right .
\end{equation}
Therefore, the case considered in the article is essentially a general case.

\section{Generalized Lorentz transformations}
Let the coordinates of the event $A$ in the inertial reference frame $S$ be equal to $(t,x)$, and the observer $S^\prime$ moves with the velocity $\beta$ relative to $S$ passing at $t=0$ the common referential event $O$. In the inertial reference frame $S^\prime$ the new time direction $t^\prime$ coincides to the direction of the world line of $S^\prime$, while the spatial axis $x^\prime$ is symmetric to $t^\prime$ with respect to the diagonal $t=x$, since Einstein's simultaneity convention ensures that the light velocity in all inertial reference frames is one.
\begin{figure}[ht]
\begin{center}
\includegraphics[width=0.4\textwidth]{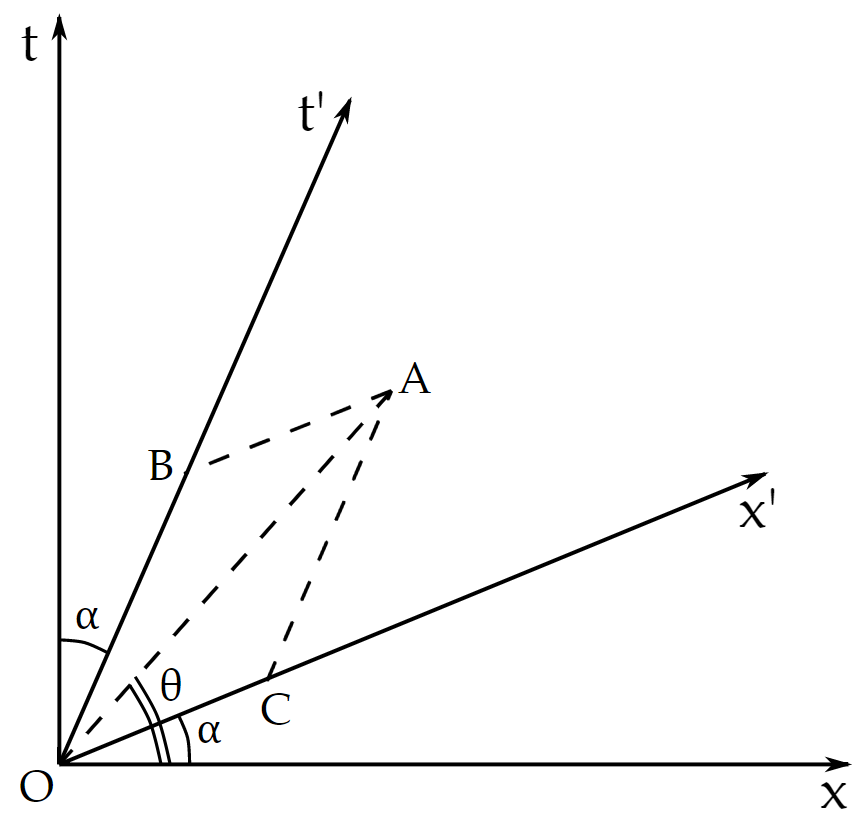}
\end{center}
\caption{Explanation of how the generalized Lorentz transformations are obtained.}
\label{fig2}
\end{figure}

\begin{figure}[ht]
\begin{center}
\includegraphics[width=0.4\textwidth]{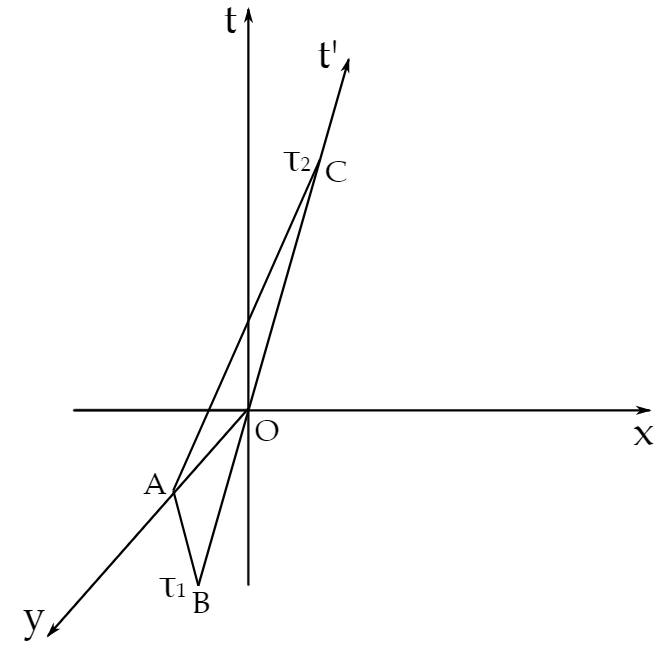}
\end{center}
\caption{Explanation of how transverse coordinate transformations are obtained.}
\label{fig3}
\end{figure}

If the inclination angle of $OA$ relative to the $x$ axis is equal to $\theta$, then $t=OA\sin{\theta}$, $x=OA\cos{\theta}$, while $t^\prime=k(\beta)\,OB$, $x^\prime=k(\beta)\,OC$ (the coefficient that converts the Euclidean length $OC$ to $x^\prime$ is the same, since the spatial distance in radar coordinates  is determined in terms of proper time). $\angle BOA=\angle OAC=\frac{\pi}{2}-(\theta+\alpha)$, $\angle AOC=\theta-\alpha$, and $\angle OCA=\pi-\angle OAC-\angle AOC=\frac{\pi}{2}+2\alpha$ (see Fig.\ref{fig2}). From the triangle $OAC$,
\begin{equation}
\frac{OA}{\sin{\left(\frac{\pi}{2}+2\alpha\right)}}= \frac{OC}{\sin{\left(\frac{\pi}{2}-(\theta+\alpha)\right)}}.
\label{eq17}
\end{equation}
Therefore,
\begin{equation}
OC=OA\,\frac{\cos{(\theta+\alpha)}}{\cos{2\alpha}}=OA\,\frac{\cos{\theta}\cos{\alpha}-\sin{\theta}\sin{\alpha}}{\cos^2{\alpha}-\sin^2{\alpha}}=\frac{\sqrt{1+\beta^2}}{1-\beta^2}\,(x-\beta t),
\label{eq18}
\end{equation}
where we have used $\tan{\alpha}=\beta$ and hence $\sin{\alpha}=\beta/\sqrt{1+\beta^2}$ and $\cos{\alpha}=1/\sqrt{1+\beta^2}$. Then
\begin{equation}
x^\prime=k(\beta)\,OC=\left (\frac{1-\beta}{1+\beta}\right )^{b/2}\frac{x-\beta t}{\sqrt{1-\beta^2}}=
\left (\frac{1-\beta}{1+\beta}\right )^{b/2}\gamma\,(x-\beta t).
\label{eq19}
\end{equation}
Similarly, from the triangle $OBA$,
\begin{equation}
\frac{OA}{\sin{\angle OBA}}=\frac{OB}{\sin{\angle BAO}}.
\label{eq20}
\end{equation}
But $\angle OBA=\angle OCA=\frac{\pi}{2}+2\alpha$ and $\angle BAO=\angle AOC=\theta-\alpha$.
Therefore,
\begin{equation}
OB=OA\,\frac{\sin{(\theta-\alpha)}}{\sin{\left (\frac{\pi}{2}+2\alpha\right )}}=OA\,\frac{\sin{\theta}\cos{\alpha}-\cos{\theta}\sin{\alpha}}{\cos^2{\alpha}-\sin^2{\alpha}}=\frac{\sqrt{1+\beta^2}}{1-\beta^2}\,(t-\beta x).
\label{eq21}
\end{equation}
Then
\begin{equation}
t^\prime=k(\beta)\,OB=\left (\frac{1-\beta}{1+\beta}\right )^{b/2}\gamma\,(t-\beta x).
\label{eq22}
\end{equation}
To find the transformation law of the transverse coordinates,  let us consider an event $A$ that lie on the $y$-axis, that is, in the $S$ reference frame it has coordinates $(0,0,y,0)$. Since $t_A=x_A=0$, the events $A$ and $O$ are simultaneous for all observers  moving along the $x$ axis. In particular, for the observer $S^\prime$ with the world line $BOC$ (see Fig.\ref{fig3}) we will have $t^\prime_A=\frac{1}{2}(\tau_1+\tau_2)=0$ and, therefore, $\tau_1=-\tau_2$, which implies $BO=OC$, since $\tau_1=-k(\beta)\,BO$, $\tau_2=k(\beta)\,OC$, where $\beta$ is the velocity of the observer $S^\prime$ in the reference frame $S$.

The Euclidean vector $\overrightarrow{AC}$ has coordinates $(t_C,\beta t_C,-y,0)$, and the angle between it and the time axis $t$ must be $\pi/4$, since $AC$ is the photon world line. Therefore,
\begin{equation}
\frac{1}{\sqrt{2}}=\cos{\frac{\pi}{4}}=\frac{\overrightarrow{AC}\cdot (1,0,0,0)}{|\overrightarrow{AC}|}=\frac{t_C}{\sqrt{t_C^2(1+\beta^2)+y^2}}.
\label{eq23}
\end{equation}
Solving with respect to $t_c$, we get $t_C=y/\sqrt{1-\beta^2}$ (the second solution corresponds to the event $B$: $t_B=-y/\sqrt{1-\beta^2}$). Therefore, 
\begin{equation}
OC=\sqrt{t_C^2(1+\beta^2)}=\sqrt{\frac{1+\beta^2}{1-\beta^2}}\,y, \;\;\;\mathrm{and}\;\;\;
\tau_2=k(\beta)\,OC=\left (\frac{1-\beta}{1+\beta}\right )^{b/2} y.
\label{eq24}
\end{equation}
In the reference frame $S^\prime$, the distance between the simultaneous events $A$ and $O$ (coordinate $y^\prime$) is equal to $y^\prime=\frac{1}{2}(\tau_2-\tau_1)=\tau_2$, and as a result we get
\begin{equation}
y^\prime=\left (\frac{1-\beta}{1+\beta}\right )^{b/2} y.
\label{eq25}
\end{equation}
As we see, the generalized Lorentz transformations have the form 
\begin{equation}
\begin{aligned}
& t^\prime=\lambda(\beta)\,\gamma\,(t-\beta x), \;\;\;\;\;
  x^\prime=\lambda(\beta)\,\gamma\,(x-\beta t), \\
& y^\prime=\lambda(\beta)\, y, \;\;\; z^\prime=\lambda(\beta)\, z, \;\;\;
\lambda(\beta)=\left (\frac{1-\beta}{1+\beta}\right )^{b/2}.
\end{aligned}
\label{eq26}
\end{equation}
These transformations leave the Finslerian interval (\ref{eq1}) invariant provided that the $x$ axis along which the observer $S^\prime$ moves is in the preferred direction $\vec{n}$. 

Interestingly, both Einstein and Poincar\'{e} obtained the $\lambda$-Lorentz transformations, but then claimed $\lambda(\beta)=1$, essentially based on spatial isotropy  \cite{Chashchina:2016tey}. Finsler space-time arises in a more general situation when spatial isotropy is not assumed, and in this case the transformations (\ref{eq26}) can be obtained in a more traditional way too, requiring the group property of $\lambda$-Lorentz transformations \cite{Chashchina:2016tey}. When the observer $S^\prime$ does not moves along the preferred direction $\vec{n}$, the generalized Lorentz transformations that leave the Finslerian interval (\ref{eq1}) invariant have more complicated form. They can be found in \cite{Bogoslovsky,Chashchina:2016tey}.

\section{Concluding remarks}
In this article, we do not touch on the question of whether Finslerian generalization of special (and general) relativity is physically viable. There is already a large amount of literature on this subject. See, for example, \cite{Finsler1,Finsler2,Finsler3,Finsler4,Finsler5} and references therein. Our aim was simply to show that the Finsler space-time geometry of a special kind, introduced in \cite{Bogoslovsky:1973zen,Bogoslovsky}, arises quite naturally as a generalization of the Robb-Geroch definition of the relativistic interval, and this generalization requires nothing more than a simple trigonometry.  

Physical implications of the Finslerian metric (\ref{eq6}) and its general-relativistic generalization was considered in detail in \cite{Bogoslovsky:1994cq,Bogoslovsky1992}. Aether wind type experiments indicate that $b<5\times 10^{-10}$ and potentially the upper bound can be lowered down to $\sim 10^{-14}$ \cite{Bogoslovsky:2007gt}. While Hughes-Drever type constraints on the anisotropy of inertia reduce the upper bound, albeit in a model-dependent manner, to fantastically small value $|b|<10^{-26}$\cite{Gibbons:2007iu}. Nevertheless, the study of the metric (\ref{eq6}) and its physical consequences is of both theoretical interest \cite{Gibbons:2007iu,Dhasmana:2019jil} and of possible importance for the physics of the early Universe \cite{Bogoslovsky:2007gt,Elbistan:2020mca}.

Fundamental assumptions justifying the use of pseudo-Riemannian geometry in relativity were investigated in the classic paper of Ehlers, Pirani, and Schild \cite{Ehlers} (for pedagogical exposition, see \cite{Linnemann:2021pea}). This highly influential work continues the tradition of basing the kinematics of relativity on a set of empirically grounded axioms about the propagation of light and free fall, a tradition that, in the case of special relativity, was started by Robb \cite{Robb1,Robb2,Robb3 ,Robb4} (nicknamed "the Euclid of relativity" \cite{Sanchez-Ron1987}). Some implicit and explicit assumptions of \cite{Ehlers} are not fully empirically motivated, and by relaxing them one can include Finsler space-times in this scheme \cite{Roxburgh,Tavakol_1985,Tavakol_1986,Finsler3}. It is interesting to note that the flat Finsler metric (\ref{eq6}) is a special case of Finsler metrics compatible with the generalized Ehlers-Pirani-Shield axiomatization considered in \cite{Roxburgh,Tavakol_1985,Tavakol_1986}, although it seems that the authors of these papers do not were aware of this special case as they do not cite Bogoslovsky's contribution \cite{Elbistan:2020mca}.

\section*{acknowledgments} 
We thank the anonymous reviewer for constructive comments that helped improve the presentation of the manuscript. The work is supported by the Ministry of Education and Science of the Russian Federation. 

\bibliographystyle{ws-mpla}
\bibliography{Finsler_SR}

\end{document}